\documentclass[slac_one]{revtex4}
\usepackage{graphicx}
\usepackage{fancyhdr}
\begin{document}

\title{ATLAS RPC offline monitoring and data quality assessment} 

%

\author{G. Chiodini\footnote{email: gabriele.chiodini@le.infn.it; 
presented at ICHEP2008, Philadelphia, USA, July 2008}}
\affiliation{INFN-Lecce, via Arnesano 73100 Lecce - Italy}
\author{M. Bianco, E. Gorini and A. Guida}
\affiliation{Universit\'{a} del Salento - Dipartimento Di Fisica and INFN-Lecce, via Arnesano 73100 Lecce - Italy}

\begin{abstract}
In this work several aspects of ATLAS RPC offline monitoring and data quality assessment
are illustrated with cosmics data selected by RPC trigger.
These correspond to trigger selection, front-end mapping, detection efficiency and occupancy,
which are studied in terms of low level quantities such as: RPC off-line hits and standalone tracks.
The tools and techniques presented are also extended to the forthcoming LHC p-p beam collisions.

\end{abstract}

\maketitle

\thispagestyle{fancy}


\section{INTRODUCTION} 

The muon spectrometer of the ATLAS experiment at the Large Hadron Collider (LHC) is built 
around three large super-conducting air-core toroids. In the barrel region, where the magnetic field is 
provided by eight radial coils, muon triggering is accomplished by 
596 Resistive Plate Chambers \cite{santonico} arranged radially at about 5 m, 7.5 m, and 10 m from the beam line \cite{muTDR}. 
The first two are located inside the toroid coils and named low-pt and pivot planes respectively;
while the outer one is located just outside the toroid and named high-pt plane.

The RPC planes are made of one or two mechanically independent RPC units (for a total of 1116). 
Each unit consists of 2 layers of active gas volume, each one instrumented with two orthogonal readout 
strip panels (measuring the bending and non-bending views with a pitch of about 3 cm) 
with built-in fast GaAs front-end electronics.
The area covered by the RPC detector is 3650 m$^{2}$ and the front-end electronics consists of approximately 
355,000 readout channels.  

The off-line monitoring and data quality assessment 
of such a large sub-system are crucial to maximize
the physics reach of the experiment. 
This can be accomplished by a detailed knowledge of the detector
performance during runs and ensuring a uniform detector behavior 
between runs in order to reduce systematic errors. 

\section{RPC DATA QUALITY}

\begin{figure}[t]
\centering
\includegraphics[width=115mm]{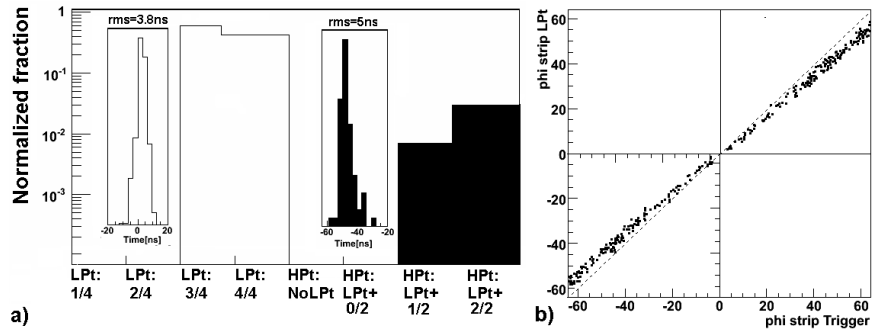}
\caption{a) RPC layers pattern of non-bending view coincidence matrix delivering a low-pt (high-pt) trigger.
In the insets the not-filled (filled) plot is the relative time distributions between pivot plane 
and low-pt (high-pt) plane.
b) Trigger road in non-bending view as measured from cosmic data. Vertical tracks correspond to the
dashed line.} 

\label{triggerconditionstiming}
\end{figure}

The readout and trigger scheme is implemented by on-detector 
programmable Coincidence Matrix (CMA) ASICs \cite{muTDR}.
A CMA trigger selection consists of a fast geometrical 25 ns temporal coincidence 
of 3 out of 4 RPC layers for low-pt triggers and 
1 out of 2 RPC layers for high-pt triggers, in addition to a low-pt trigger.    
Figure \ref{triggerconditionstiming}.a) shows the RPC layers pattern when
a trigger is present for cosmics data.
The trigger window could be further decreased, in steps of 3.125 ns, 
thanks to the excellent 1 ns timing  resolution of the RPC detector. 
The two inset plots of Figure \ref{triggerconditionstiming}.a) illustrate the relative time between 
RPC planes belonging to the same trigger tower. 
Inside a tower, the measured spread in time is dominated by the signal propagation speed along
the detecting strip; instead between towers, it is dominated by the particle time-of-flight.
Both effect can be corrected off-line and the RPC timing is a powerful tool against cosmics, 
cavern background, and events pile-up.
The width of the geometrical coincidence (named `trigger road') can be 
programmed up to 3+3 high transverse momentum track values.
Figure \ref{triggerconditionstiming}.b) clearly shows a projective trigger road 
in non-bending view as extracted from cosmics data.

The RPC mapping is a not trivial
task because the same electronics implements
the trigger logic and the readout.
In fact, to avoid trigger inefficiency a large fraction of RPC strips
are readout by two adjacent coincidence matrix in the 
low-pt and high-pt planes (named `cabling overlaps').
The pointing geometry requires cabling overlaps which are position dependent
along the beam and when chamber boundaries are crossed in the bending view,
a full non-bending view overlap is required between chambers (named `logical-or').

\begin{figure}[t]
\centering
\includegraphics[width=115mm]{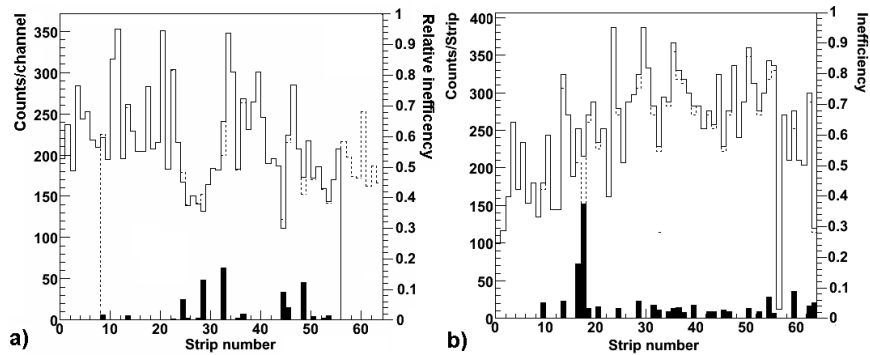}
\caption{ a) The full and dashed lines represent the channel profiles of two coincidence matrix 
reading out a non-bending view 64-strip detector panel. The filled plot 
is the relative counting difference in the overlap region.
b) The full and dashed lines represent the strip profiles of projected tracks and corresponding efficient strips. 
The filled plot is the single strip inefficiency.}\label{cablingoverlapefficiency}
\end{figure}

Figure \ref{cablingoverlapefficiency}.a) shows an example of channel overlaps between
two coincidence matrix, together with the relative inefficiency obtained 
with cosmic rays.
The resulting overlap inefficiency is a fraction of percent and compatible
with the random arrival time of cosmics inside a 25 ns clock period.

The software tools should be capable 
to produce a fast feedback on RPC detector data quality,
without relay on the full ATLAS event 
reconstruction and combined quantities.
A RPC standalone tracking is implemented
in off-line monitoring  \cite{chiodini}.
The tracking is based on RPC space points, which are defined
by orthogonal RPC cluster hits of the same
gas volume.
Figure \ref{clustersizeoccupancy}.a) shows the average readout panel cluster size distribution
at a gas volume high voltage value of 9600 V and nominal front-end voltage threshold.

The pattern recognition is seeded 
by a straight line, which is defined by 
two RPC space points belonging, respectively,
to low-pt and pivot planes of the same or nearby station.
RPC space points not part of any previous tracks
and inside a predefined distance from the straight line are associated
to the pattern.
Resulting patterns with points in at least 3 out of 4 layers in low-pt and pivot
planes are retained and a linear interpolation is performed in two orthogonal views.
From cosmic data about 95 \% percent of events have at least one RPC track;
this is due the strong correlation between the pattern recognition and the trigger algorithm.
Applying a quality cut of chi2/dof $<$ 1 about 70 \% of events have at least a good tracks
and 10 \% with more than one. 

In order to measure the detection efficiency the RPC tracking
is repeated 6 times, each time removing the layer under test from the pattern recognition and track fitting.
Figure \ref{cablingoverlapefficiency}.b) shows the strip profile of readout panel under test
in correspondence of the projected track and the profile of efficient strip. 
The strip efficiency is also shown on the same plot and it results in an
unbiased measure because of the 3 out of 4 majority trigger logic.

\section{COSMIC RAYS VERSUS COLLISIONS}

The RPC tracking standalone is extended to the forthcoming LHC p-p beam collisions.
Cosmic rays arrive randomly in time and not uniformly on detector 
surface. This makes detector studies with cosmics not very accurate and predictable.   
Tracks produced by beam collisions are synchronous with beam clock,
pointing, and uniform in azimuthal angle and pseudo-rapidity.
The difficultly with beam is due to the presence of the magnetic field
and operation at high luminosity.
The above described pattern recognition and the
track quality cut correspond, in magnetic field,
to a cut in transverse momentum.
At high luminosity a large uncorrelated and correlated background could increase
the number of fake tracks significantly. 
In order to mitigate such a problem we tune the pattern recognition
and the track quality cut and reject low momentum tracks.

\begin{figure}[t]
\centering
\includegraphics[width=100mm]{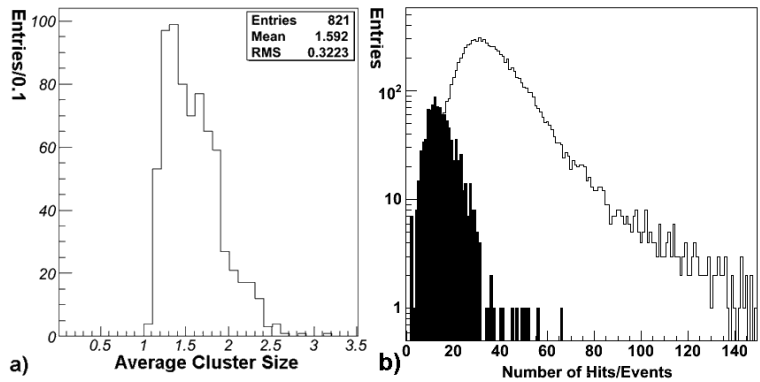}
\caption{ a) Clusters size distribution of RPC readout strips.
b) Hits multiplicity in the event considering cosmic rays triggered by RPC (non filled plot) and
random trigger (filled plot).} \label{clustersizeoccupancy}
\end{figure}

In Figure \ref{clustersizeoccupancy} the distribution of RPC hits per event 
with RPC trigger and with random trigger are shown.
The average value of these distributions correspond to the average RPC detector
occupancy due to cosmics and uncorrelated noise.
The cosmic ray and random trigger data show a low level of uncorrelated noise 
in RPC detector, which corresponds
to fraction of Hz$/$cm$^2$. 
During beam collisions, at nominal luminosity, 
uncorrelated and correlated noise is going to be dominated
by cavern background and low energy tracks.

\section{CONCLUSIONS}
Data quality on trigger selection and detector performance
could be asses in a quite straight forward and simple
way by looking to distributions of layers pattern in trigger, single channel overlap
efficiency, single strip efficiency, and average readout panel occupancy.
The monitoring of these distributions guarantees good data on tape and promptly spot eventually occurring problems.

\end{document}